\documentclass[letterpaper,twocolumn,dvipdfmx]{jpsj3}
\usepackage{txfonts}
\usepackage{bm}
\usepackage{color}
\usepackage{comment}

\title{Symmetry lowering on the field-induced commensurate phase in CeRhIn$_5$}

\author{Tomoki Kanda $^1$\thanks{t.kanda@issp.u-tokyo.ac.jp}, Koki Arashima$^2$, Yusuke Hirose$^3$, Rikio Settai$^3$, Kazuki Matsui$^1$, Toshihiro Nomura$^1$, Yoshimitsu Kohama$^1$ and Yoshihiko Ihara$^2$\thanks{yihara@phys.sci.hokudai.ac.jp}}
\inst{$^1$The Institute of Solid State Physics, University of Tokyo, Kashiwa, Chiba 277-8581, Japan \\
$^2$Department of Physics, Faculty of Science, Hokkaido University, Sapporo 060-0810, Japan \\
$^3$Graduate School of Science and Technology, Niigata University, Niigata 950-2181, Japan }

\abst{Temperature dependence of the $^{115}$In-NMR spectra of CeRhIn$_5$ is studied with the external magnetic fields 10$^\circ$ off the [100] and [001] axes. Our detailed analyses confirm that the AFM3 phase breaks the four-fold spin symmetry with the commensurate ordering vector of $\bm{Q} = (0.5, 0.5,0.25)$. Based on the observation of anistropic hyperfine fields, we also propose the symmetry lowering of the electronic structure in the AFM3 phase.}

\begin{document}
\maketitle

\section{Introduction}
\begin{figure}
\begin{center}
\includegraphics[width=70mm]{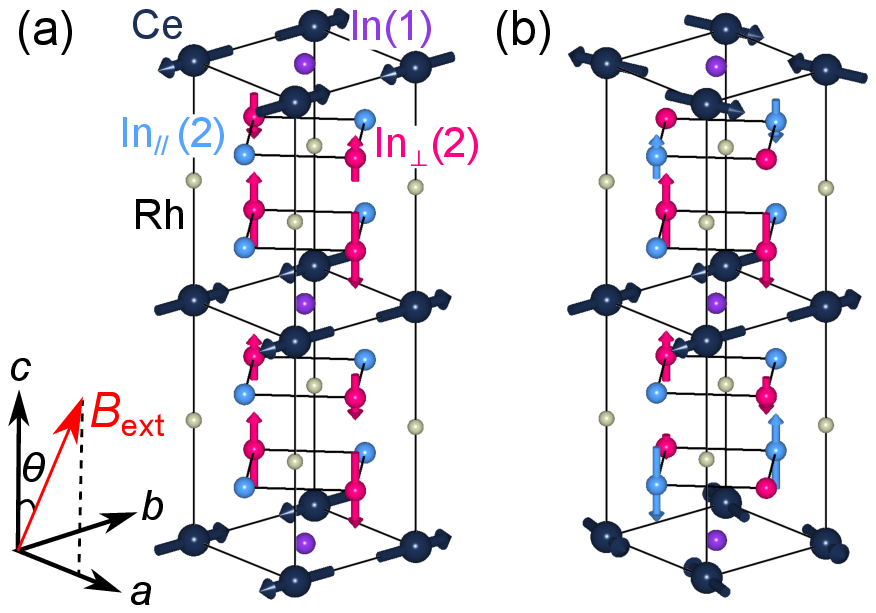}
\caption{Crystal and magnetic structures of CeRhIn$_5$. (a) Commensurate AFM3 ($\bm{Q} = (0.5,0.5,0.25)$) and (b) incommensurate helical AFM1 ($\bm{Q} = (0.5,0.5,0.294)$) structures suggested by the earlier works \cite{Raymond2007, Fobes2018}. The arrows on the Ce site indicate the direction of the Ce $4f$ moments projected on the (001) plane. The arrows on the In sites correspond to the calculated internal field.
}
\label{f1}
\end{center}
\end{figure}

Ce-based heavy-fermion system has strong competition between RKKY interaction and Kondo effect, providing a fertile ground for studying novel phenomena. \cite{Lohneysen2007, Philipp2008}
A prototypical system is CeRhIn$_5$ which crystallizes in a tetragonal structure $(P4/mmm)$ with alternating stacks of antiferromagnetic (AFM) CeIn$_3$ and non-magnetic RhIn$_2$ layers (Fig. \ref{f1}) \cite{Moshopoulou2001}.
After the report of the $d$-wave superconductivity above 1.0 GPa, \cite{Park2008} extensive high-pressure experiments have been devoted to unveil the relationship between the antiferromagnetism and the superconductivity \cite{Yashima2009, Park2008, Park2012}.
Recently, an electronic nematic state is found in CeRhIn$_5$ when the external field of $\sim$30 T is applied with the tilting angle of 20$^\circ$ from the [001] axis \cite{Ronning2017}.
In this state, magnetoresistance becomes anisotropic for $[100]$ ($[110]$) and $[010]$ ($[\overline{1}10]$) which are supposed to be equivalent in tetragonal crystal symmetry.
Since the sign of the anisotropic magnetoresistance can be inverted with the external field direction, the electronic nematic state in CeRhIn$_5$ is proposed to be different from the ordinary anisotropy by the crystal symmetry breaking and regarded as an XY nematic state \cite{Ronning2017}.

Figure \ref{f1} shows crystal and magnetic structures of CeRhIn$_5$.
The crystallographic unit cell indicated by the gray box contains Ce and Rh atoms at each corner and the edge of the [001] axis, respectively.
In(1) sites locate on the (001) plane and In(2) sites locate on the (100) or (010) plane.
Under in-plane magnetic fields, the In(2) becomes two inequivalent sites due to the electric field gradient (EFG).
For $\bm{B}_\mathrm{ext} \parallel [100]$, we distinguish the two sites located in the  (100) and (010) planes as In$_\perp$(2) and In$_\parallel$(2) sites, respectively.

Below $T_\mathrm{N}$ = 3.8 K in zero field, AFM1 phase appears (Fig. \ref{f1}(b)), which is characterized by the incommensurate helical ordering of the Ce $4f$ moments along the wavevector of $\bm{Q} = (0.5, 0.5, 0.294)$ \cite{Bao2000}.
The AFM1 phase is stable when the external magnetic field is applied perfectly parallel to the [001] axis up to the critical field of $\sim$50 T.
When the external magnetic fields are applied perpendicular to the [001] axis, in contrast, a metamagnetic transition takes place at $B_\mathrm{MM}=2.1$~T from AFM1 to AFM3 phase \cite{Cornelius2000, Takeuchi2001, Jiao2015}.
The AFM3 phase is characterized by the commensurate collinear antiferromagnetic order with $\bm{Q} = (0.5,0.5,0.25)$ as shown in Fig. \ref{f1}(a), which is called “up-up-down-down” configuration and breaks four-fold rotational symmetry of crystalline lattice.
Here, the Ce moments are proposed to align perpendicular to the magnetic field \cite{Raymond2007, Fobes2018}.

The AFM3 phase is believed to appear when the projection of the tilted magnetic fields on the (001) plane is higher than $B_\mathrm{MM}$. \cite{Rosa2019}
D. M. Fobes \textit{et al.} suggest that the magnetic symmetry breaking in the AFM3 phase might relate to the appearance of the electronic nematic state \cite{Fobes2018}.
On the other hand, the quantum oscillation and high field NMR measurements with the applied fields parallel to the [001] axis also detect the change of electronic structure at $\sim$30 T. \cite{Jiao2019, Lesseux2019}.
These results suggest that the electronic nematic state can be realized without entering the AFM3 phase.
The relationship between the AFM3 phase and the electronic nematic state is still controversial.

In this paper, we present the $^{115}$In-NMR spectra of CeRhIn$_5$ in the magnetic field slightly tilted from the [100] and [001] axes.
The temperature dependence of the NMR spectra clearly indicates the symmetry lowering of the internal fields due to the magnetic ordering.
The fine structures of the NMR spectra are discussed in terms of the dipole and hyperfine interactions.
Combined with the simulation of the internal fields, the magnetic structure and the site symmetry of CeRhIn$_5$ are discussed.
We also discuss how the magnetic ordering and the concomitant symmetry braking relate to the electronic nematicity.

\section{Experiment}
Single crystals of CeRhIn$_5$ were grown by the self-flux method.\cite{Shishido2002}
No In inclusions originating from the flux were detected in the present NMR data.
The size of the sample used for this study was $0.5 \times 0.5 \times 1 \; \mathrm{mm}^3$. The longest axis was parallel to the crystalline [010] direction, which was parallel to the rotation axis of our single axis rotator.

The field-sweep NMR spectra at the fixed frequency of 123.51~MHz were obtained by recording the Fourier transformation of the spin-echo signals during the field sweeps.
The field orientation with respect to the crystallographic axes was determined from the peak positions of the field-sweep NMR spectra.
In this study, we chose the angles $\theta \sim 10^\circ$ and $80^\circ$, where $\theta$ is the polar angle between $B_\mathrm{ext}$ and the [001] axis (Fig. \ref{f1}).
The azimuthal angle was close to zero, namely, the external field projected to the (001) plane was parallel to the [100] axis.

\section{Results and Discussion}
\begin{figure}
\begin{center}
\includegraphics[width=70mm]{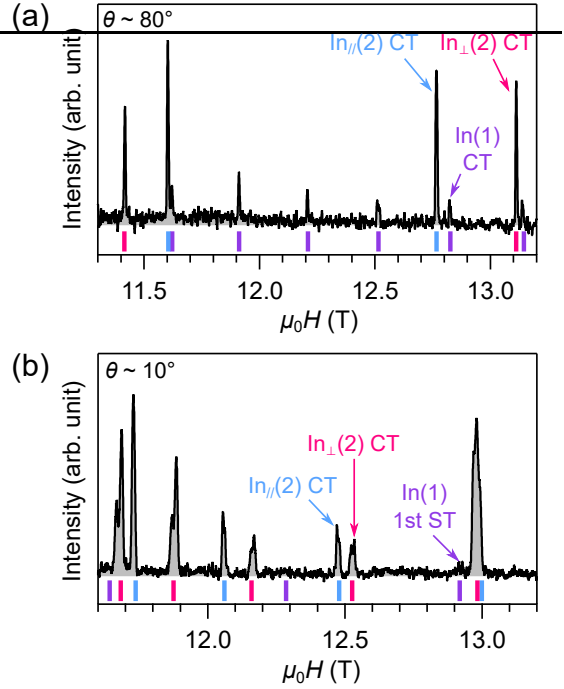}
\caption{NMR spectra of CeRhIn$_5$ for (a) $\theta \sim 80^\circ$ and (b) $\theta \sim 10^\circ$ in the paramagnetic phase at 6.0 K. These spectra were obtained at 123.51 MHz. The short vertical bars at the bottom indicate the calculated peak positions based on Eq. (\ref{eq1}). The purple, red, and blue bars represent the contributions from In(1), In$_\perp$(2), and In$_\parallel$(2) sites, respectively.}
\label{f2}
\end{center}
\end{figure}

Figure \ref{f2} shows the $^{115}$In-NMR spectra in the paramagnetic (PM) phase of CeRhIn$_5$ at 6.0~K.
NMR signals from three In sites, In(1), In$_\parallel$(2), and In$_\perp$(2), are observed at different fields.
Since the in-plane component of the external field is along [100] axis, In$_\perp$(2) and In$_\parallel$(2) correspond to the In sites on the (100) and (010) planes, respectively.
Each $^{115}$In-NMR peak splits into 9 peaks (with nuclear spin $I = 9/2$) due to the nuclear-quadrupole interaction, resulting in 27 NMR peaks in the PM phase. Within our field sweep range, the central transition (CT) of $+1/2 \Leftrightarrow -1/2$ was mostly observed.

The peak positions in Fig. 2 can be reproduced by the following nuclear spin Hamiltonian,
\begin{equation}
\label{eq1}
\mathcal{H} = \hbar \gamma (1+K) \bm{I} \cdot \bm{B} +  \frac{h\nu_Q}{6} \left[ 3{I_z}^2-\bm{I}^2 + \eta({I_x}^2 - {I_y}^2 ) \right].
\end{equation}
Here, $K$ is the Knight shift, $\nu_Q$ is the quadrupolar frequency, and $\eta$ is the asymmetric parameter of EFG.
The first and second terms in Eq. (\ref{eq1}) represent the Zeeman energy and the nuclear-quadrupole interaction, respectively.
We set the parameters suggested by the previous work as $\nu_Q=6.78$ MHz and $\eta=0.0$ for In(1), $\nu_Q=16.665$ MHz and $\eta=0.445$ for both In$_\perp$(2) and In$_\parallel$(2) \cite{Curro2000}.
The peak positions of the simulated spectrum are indicated by the vertical bars at the bottom of Figs. 2(a) and 2(b).
The purple, red, and blue bars indicate the contributions from In(1), In$_\perp$(2), and In$_\parallel$(2) sites, respectively.
The best fits are obtained with the parameters $\theta=79^\circ$, $K(\mathrm{In(1)}) = 2.9 \%$, $K(\mathrm{In_\perp(2)}) = 1.35 \%$, and $K(\mathrm{In_\parallel(2)}) = 1.85 \%$ in Fig. 2(a), and with $\theta = 10^\circ$ and $K(\mathrm{In(1)}) = 7.9 \%$, $K(\mathrm{In_\perp(2)}) = 2.45 \%$ and $K(\mathrm{In_\parallel(2)}) = 2.5 \%$ in Fig. 2(b).
The obtained $K$ agrees well with earlier reports \cite{Curro2003, Lin2015}.
We note that the first satellite (ST) $(1/2 \Leftrightarrow 3/2)$ of In(1) was used for the fit in Fig. \ref{f2}(b) because the peak corresponding to the central transition is smeared out.
The obtained values of $\theta$ corroborate the field directions independently determined by the single-axis rotator, and the error of $\theta$ is estimated to be $\pm 1^\circ$.

\begin{figure}
\begin{center}
\includegraphics[width=70mm]{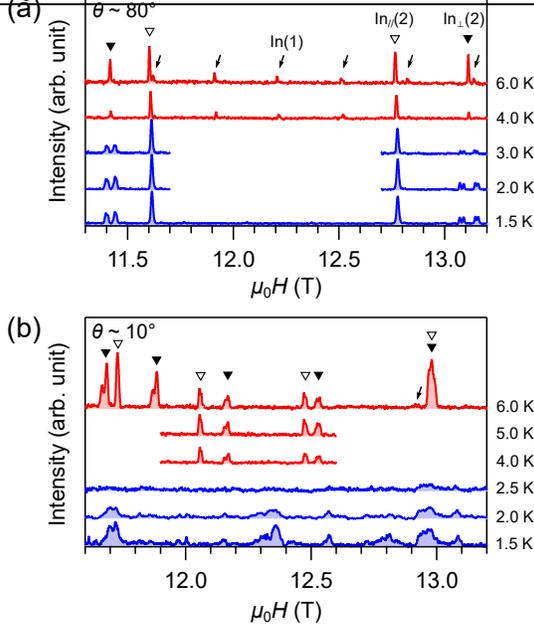}
\caption{Temperature dependence of the NMR spectra for (a) $\theta \sim 80^\circ$ and (b) $\theta \sim 10^\circ$. The arrows, filled triangles, and open triangles represent the peaks from In(1), In$_\perp$ (2), and In$_\parallel$ (2) sites, respectively.}
\label{f3}
\end{center}
\end{figure}

Figure \ref{f3} shows the temperature dependence of the NMR spectra with the peak assignment from Fig. 2.
At low temperatures below $T_\mathrm{N}$, most of the NMR peaks split by the magnetic dipole and hyperfine interactions from the ordered moments.
Particularly, in Fig. 3(a), In$_\perp$(2) peaks split, while In$_\parallel$(2) peaks do not.
This result suggests that the magnetic ordering creates two inequivalent In$_\perp$(2) sites by breaking symmetry, however, the effect is canceled at In$_\parallel$(2) sites.
The difference is well explained by the suggested magnetic structure of AFM3 in Fig.~1(a) as discussed later.

In Fig. \ref{f3}(a), the line splittings of In$_\perp$(2) are observed at two different magnetic fields of 11.41 and 13.11 T.
The former corresponds to $-1/2 \Leftrightarrow -3/2$, and the latter corresponds to the central transition of $+1/2 \Leftrightarrow -1/2$.
Since the central transition is not affected by the quadrupole interaction in the first order perturbation to Eq. (\ref{eq1}), we focus on the central transition in the following discussions.

\begin{figure}
\begin{center}
\includegraphics[width=70mm]{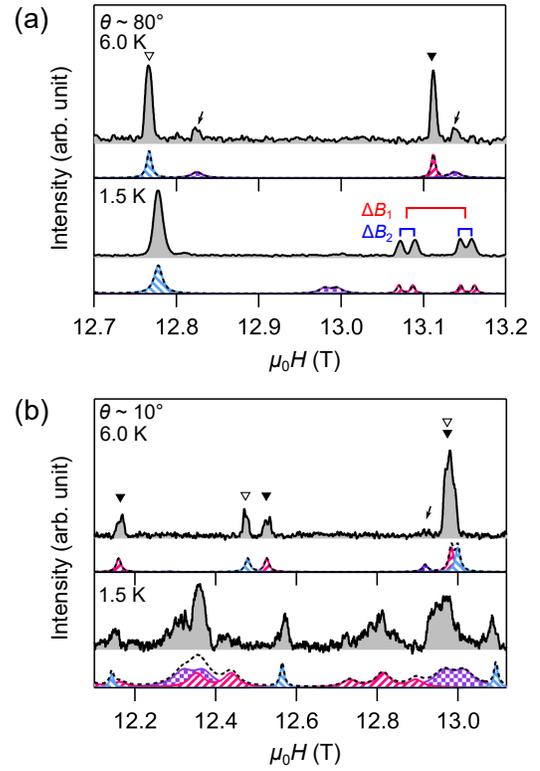}
\caption{NMR spectra near the central transitions for (a) $\theta \sim 80^\circ$ and (b) $\theta \sim 10^\circ$.
The simulated NMR spectra are shown by the purple (In(1)), red (In$_\perp$(2)), and blue (In$_\parallel$(2)) shadowed area in the bottom of the plot. The dotted line shows the summation.}
\label{f4}
\end{center}
\end{figure}

Figure \ref{f4} shows the enlarged NMR spectra around the central transition in the PM phase and the magnetic ordered state at 1.5 K.
In the case of $\theta \sim 80^\circ$ [Fig. \ref{f4}(a)], the peak of In$_\perp$(2) at 13.11 T splits into two groups with a large separation of 70 mT ($\Delta B_1$).
Each group further splits into two peaks with a small separation of 15 mT ($\Delta B_2$), leading to the four-peak structure in the ordered state.
Note that the four-peak structure can also be interpreted as two discrete splitting of $\Delta B_1 - \Delta B_2$ and $\Delta B_1+ \Delta B_2$.
On the other hand, the NMR peaks for In$_\parallel$(2) do not show any splitting.
In the case of $\theta \sim 10^\circ$ [Fig. \ref{f4}(b)], most NMR lines broaden and overlap with neighboring peaks.

Here, we discuss the magnetic structure in the ordered phase based on the shape of the NMR spectra in Fig.~\ref{f4}.
If the magnetic order is incommensurate, spatially modulated internal magnetic fields at each In site result in a peculiar ''double horn'' pattern \cite{Curro2000, Yashima2009}.
In contrast, sharp NMR lines should remain in the commensurate phase where the nuclei feel the distinct values of the internal magnetic field ($\bm{B}_\mathrm{int}$).
Therefore, NMR lines for the incommensurate AFM1 should become broader than those of the PM phase \cite{Curro2000, Yashima2009}.
The shape of the NMR spectra in Fig. \ref{f4}(a) ($\theta \sim 80^\circ$) does not show significant broadening, and is consistently explained by the commensurate AFM3 structure.

With assuming the AFM3 structure, we explain the origin of $\Delta B_1$ and $\Delta B_2$. In general, $\bm{B}_\mathrm{int}$ is caused by the dipolar magnetic field from magnetic moments ($\bm{B}_\mathrm{dip}$) and the hyperfine magnetic field ($\bm{B}_\mathrm{hyp}$).
We calculate $\bm{B}_\mathrm{dip}$ by the following equation based on the classical electromagnetism:
\begin{equation}
\label{eq2}
\bm{B}_\mathrm{dip} = \sum_{\mathrm{Ce \; sites}} -\frac{\mu_0}{4\pi} \left[ \frac{\bm{m}}{r^3} - \frac{3(\bm{m \cdot r})\bm{r}}{r^5} \right].
\end{equation}
Here, $\bm{r}$ is the position vector from the In to the Ce sites and $\bm{m}$ is the $4f$ magnetic moment. In CeRhIn$_5$, $\bm{m}$ for the AFM3 structure was proposed as \cite{Raymond2007}
\begin{equation}
\label{eq3}
\bm{m}_i = 0.59\mu _\mathrm{B} \cdot \sqrt{2} \cos \left(\frac{\pi x_i}{a} \right) \cos \left( \frac{\pi y_i}{a} \right) \sin \left( \frac{\pi z_i}{2c} + \frac{\pi}{4} \right) \bm{\hat{y}}.
\end{equation}
Here, we introduce the unit cell coordinate system, where $\bm{\hat{x}}, \bm{\hat{y}}, \bm{\hat{z}}$ are along the [100], [010], [001] axes respectively. $0.59\mu_\mathrm{B}$ is the size of Ce moment \cite{Raymond2007}, $a = 4.656$ \AA \ and $c = 7.542$ \AA \ are the lattice constants \cite{Moshopoulou2001}, and $\bm{r}_i=(x_i,y_i,z_i)$ is the coordinate of the $i$-th Ce sites.
In addition to the ordered moments, we take the induced moments into consideration.
To a first approximation, we treat the induced moment parallel to $\bm{B}_\mathrm{ext}$ as,
\begin{equation}
  \label{eq3.1}
  \delta \bm{m} = \delta m \frac{\bm{B}_\mathrm{ext}}{|\bm{B}_\mathrm{ext}|} = \delta m (\bm{\hat{x}} \sin \theta + \bm{\hat{z}} \cos \theta ).
\end{equation}
Thus, the Ce moments under magnetic fields are described as ${\bm{m}_i}^{\prime} = \bm{m}_i + \delta \bm{m}$.

$\bm{B}_\mathrm{hyp}$ originates from the on-site hyperfine interaction and the transferred hyperfine interaction. Since these are the indirect interaction between the nuclear and the ordered moments through the conduction electron, the hyperfine magnetic field reflects the electronic structure.
In the AFM3 phase, the on-site hyperfine interaction is zero for the antiferromagnetic structure, while the transferred hyperfine field at In sites is calculated by the following equations \cite{Curro2006},
\begin{eqnarray}
  \label{eq4.0}
  \bm{B}_\mathrm{hyp}\left( \mathrm{In(1)} \right) &=& 2 \left(
  \begin{array}{ccc}
  {B_0}^{\prime} & {B_a}^{\prime} & 0 \\
  {B_a}^{\prime} & {B_0}^{\prime} & 0 \\
  0 & 0 & 0
  \end{array}
  \right) \left(
  \begin{array}{ccc}
  \delta m \sin \theta \\
  m \\
  \delta m \cos \theta
  \end{array}
  \right) \nonumber \\
  &+& 2 \left(
  \begin{array}{ccc}
  {B_0}^{\prime} & -{B_a}^{\prime} & 0 \\
  -{B_a}^{\prime} & {B_0}^{\prime} & 0 \\
  0 & 0 & 0
  \end{array}
  \right) \left(
  \begin{array}{ccc}
  \delta m \sin \theta \\
  -m \\
  \delta m \cos \theta
  \end{array}
  \right) \nonumber \\
  &=& 4{B_a}^{\prime} \left(
  \begin{array}{ccc}
  m \\
  0 \\
  0
  \end{array}
  \right) + 4{B_0}^{\prime} \delta m \left(
  \begin{array}{ccc}
  \sin \theta \\
  0 \\
  0
  \end{array}
  \right),
\end{eqnarray}

\begin{eqnarray}
\label{eq4}
\bm{B}_\mathrm{hyp}\left( \mathrm{In_\parallel (2)} \right) &=& \left(
\begin{array}{ccc}
B_0 & 0 & B_a \\
0 & B_y & 0 \\
B_a & 0 & B_0
\end{array}
\right) \left(
\begin{array}{ccc}
\delta m \sin \theta \\
m \\
\delta m \cos \theta
\end{array}
\right) \nonumber \\
&+& \left(
\begin{array}{ccc}
B_0 & 0 & -B_a \\
0 & B_y & 0 \\
-B_a & 0 & B_0
\end{array}
\right) \left(
\begin{array}{ccc}
\delta m \sin \theta \\
-m \\
\delta m \cos \theta
\end{array}
\right) \nonumber \\
&=&
2B_0 \delta m \left(
\begin{array}{ccc}
\sin \theta \\
0 \\
\cos \theta
\end{array}
\right),
\end{eqnarray}

\begin{eqnarray}
\label{eq5}
\bm{B}_\mathrm{hyp}\left( \mathrm{In_\perp (2)} \right) &=& \left(
\begin{array}{ccc}
B_y & 0 & 0 \\
0 & B_0 & B_a \\
0 & B_a & B_0
\end{array}
\right) \left(
\begin{array}{ccc}
\delta m \sin \theta \\
m \\
\delta m \cos \theta
\end{array}
\right) \nonumber \\
&+& \left(
\begin{array}{ccc}
B_y & 0 & 0 \\
0 & B_0 & -B_a \\
0 & -B_a & B_0
\end{array}
\right) \left(
\begin{array}{ccc}
\delta m \sin \theta \\
-m \\
\delta m \cos \theta
\end{array}
\right) \nonumber \\
&=& 2B_a \left(
\begin{array}{ccc}
0 \\
0 \\
m
\end{array}
\right) + 2 \delta m \left(
\begin{array}{ccc}
B_y \sin \theta \\
0 \\
B_0 \cos \theta
\end{array}
\right).
\end{eqnarray}
Here, $B_i$ and ${B_i}^{\prime}$ ($i = 0, a, y$) are constant values and $m$ is the magnitude of the Ce dipolar moment.
Although these tensor elements have not been precisely determined, the hyperfine coupling strength in the PM phase indicates that $B_i$ and ${B_i}^{\prime}$ are $\sim$0.1 T$ / \mu_\mathrm{B}$. \cite{Lin2015}
Based on Eqs. (\ref{eq4.0})-(\ref{eq5}), the effect of $\bm{m}$ and $\delta \bm{m}$ can be considered separately.
The induced moments on each Ce site equally shift all NMR lines and do not contribute to the line split.
The observed line splittings are discussed as a consequence of the ordered moments $\bm{m}$.

Based on these equations, the internal magnetic field in the AFM3 phase is quantitatively discussed.
As seen in Eq. (3) and Fig. \ref{f1}(a), $\bm{m}$ are aligned anti-parallel to each other and perpendicular to the external magnetic field.
For this case, $\bm{B}_\mathrm{dip}$ at In$_\parallel$(2) sites is canceled, while the finite value of $\bm{B}_\mathrm{dip}$ parallel to the [001] axis remains at In$_\perp$(2) site.
The resultant dipole magnetic fields at In$_\perp$(2) sites are calculated as $\bm{B}_\mathrm{dip} = \pm 43$ mT.
Here, we included the dipole fields from Ce moments within the distance of 100
\AA.
For $\theta \sim 80^\circ$, the NMR peak splittings are induced by the projection of the internal magnetic field to the external magnetic field as $B_\mathrm{dip} \cos 80^\circ = \pm 8$ mT, which agrees quite well with the observed splitting $\Delta B_2 /2=7.5$~mT.

Since $\bm{B}_\mathrm{dip}$ induces the small splitting $\Delta B_2$, $\bm{B}_\mathrm{hyp}$ is considered to be the origin of the large splitting $\Delta B_1$.
Although there is no report on the hyperfine magnetic field in the AFM3, we can reasonably use the value of the AFM1 phase, $B_\mathrm{int} = 250 \; \mathrm{mT}$, obtained by the previous NQR data \cite{Yashima2009}, because Eqs. (4) and (5) give the similar value of $\bm{B}_\mathrm{hyp}$ for these magnetic structures.
The $\bm{B}_\mathrm{dip}$ and $\bm{B}_\mathrm{hyp}$ are calculated to be parallel along the [001] axis, thus,
$B_\mathrm{hyp} = B_\mathrm{int} - B_\mathrm{dip} = 207$ mT is obtained.
Since the NMR peak shift depends on the projection of the internal field to the external magnetic field, $\bm{B}_\mathrm{hyp}$ induces the shift of $B_\mathrm{hyp} \cos 80^\circ = \pm 35.9$~mT, which corresponds to the observed splitting of $\Delta B_1/2=35$~mT.
Therefore we conclude that $\Delta B_1$ results from $\bm{B}_\mathrm{hyp}$.
The calculated internal fields at each In site are shown by the small arrows in Fig. \ref{f1}.
Note that the double-peak structure around 11.42 T originating from the 1st satellite transition can also be reproduced by substituting $\bm{B} = \bm{B}_\mathrm{ext} + (0, 0, \pm |B_\mathrm{hyp} \pm B_\mathrm{dip}| )$ to Eq.~(\ref{eq1}).

The NMR spectra on the In(1) sites are calculated as purple shadowed area in Fig. \ref{f4}.
Since $\bm{B}_\mathrm{dip}$ and $\bm{B}_\mathrm{hyp}$ are parallel to the [100] axis and point to the same direction, the NMR lines of In(1) split into two in the AFM3 phase.
Unfortunately, the observed In(1) peak in Fig. 4(a) was too weak for quantitative discussion.

As seen in Fig. \ref{f4}(b) ($\theta \sim 10^\circ$), the line widths of the peaks at 12.57 and 13.08 T are one third of those of the neighboring peaks.
Since both In$_\perp$(2) and In$_\parallel$(2) peaks should broaden in the AFM1 phase \cite{Curro2000, Yashima2009}, the relatively narrow peaks suggest that the AFM3 phase is stable for this field and angle range.
Indeed, with assuming AFM3 structure and taking $\bm{B} = \bm{B}_\mathrm{ext} + \bm{B}_\mathrm{int}$ in Eq. (\ref{eq1}), the NMR spectra are reproduced as the shadowed area in Fig. \ref{f4}(b).
It is found that the peak at 12.14 T is also assigned for In$_\parallel$(2), which suggests that the metamagnetic transition to the AFM3 phase already occurs below $B_\mathrm{ext} = 12.14 \; \mathrm{T}$ at $\theta \sim 10^\circ$.
With this condition, the projection of the $\bm{B}_\mathrm{ext}$ to the (001) plane is $B_{\mathrm{ext}} \sin 10^\circ = 2.11 \; \mathrm{T}$ that is compatible with  $B_\mathrm{MM} $.
Therefore, it is reasonable to observe the AFM3 at this condition.
We also note that the spectrum width for In$_\perp$(2) is broader for $\theta \sim 10^\circ$ than that for $\theta \sim 80^\circ$.
One possible explanation is that the phase boundary between AFM1 and AFM3 is located close to 12 T at $\theta \sim 10^\circ$, which can result in the large spin fluctuation and short spin-spin relaxation time ($T_2$).
The measurement of NMR spectrum with short $T_2$ is technically challenging due to the broadening of the NMR lines.

We compare our results with several recent studies on the rotational symmetry breaking of CeRhIn$_5$.
In the pioneering work by Ronning \textit{et al}. \cite{Ronning2017}, the angle dependence of the magnetoresistance reveals that the application of the tilted strong magnetic fields above 30 T break the rotational symmetry of the electronic structure.
Although the prominent anisotropic resistivity is observed only above $\sim$30 T, there is non-negligible anisotropic component in the low field AFM3 phase as well.
More recent works by inelastic neutron scattering  \cite{Fobes2018}, magnetostriction measurements \cite{Rosa2019} and ultrasonic measurements \cite{Kurihara2020} have pointed that the symmetries of magnetic and crystal structures are also broken in the AFM3 phase.
In addition to these previous studies, we experimentally found the symmetry lowering of the internal fields from the $C_4$ symmetry in the AFM3 phase, originating from its peculiar magnetic and electronic structures.

In the previous studies on the AFM1 and PM phases,\cite{Curro2000, Yashima2009, Lin2015, Lesseux2019} the magnetic field was applied to the [001] axis. 
This makes In$_\perp$(2) and In$_\parallel$(2) sites indistinguishable. 
In the present work, we applied the external magnetic field 10$^\circ$ off the crystallographic axes to enter the AFM3 phase, where  the line splittings of In$_\perp$(2) was observed, indicative of the symmetry lowering of the internal fields. 
Therefore, the occurrence of the AFM3 phase is likely a necessary condition for lowering the rotational symmetry of the internal fields, inferring that the symmetry breaking is a generic property for the AFM3 phase of CeRhIn$_5$.
Although the detailed azimuthal angle dependence of the NMR spectra is not investigated in the present study, we emphasize that the experimental NMR spectra cannot be reproduced without taking the assumption proposed by Raymond \textit{et al.} \cite{Raymond2007} and Fobes \textit{et al.} \cite{Fobes2018}, in which the Ce $4f$ moments align perpendicular to the external field direction regardless the crystallographic axes.
This indicates that the four-fold symmetry breaking of the spin and the internal fields can be switched by rotating the magnetic field direction in the same manner as the anisotropic resistivity is switched in the electronic nematic state.\cite{Ronning2017}
Therefore, we speculate that the $C_4$ symmetry breaking of the AFM3 structure may lead to the electronic nematic state observed in high magnetic fields.

\section{Summary}
We have performed the $^{115}$In-NMR spectroscopy on the heavy-fermion antiferromagnet, CeRhIn$_5$, when the external magnetic fields are $10^\circ$ off the [100] and [001] axes.
The NMR lines at In$_\perp$(2) site splits into 4 small peaks in the AFM3 phase, but not the NMR line at In$_\parallel$(2).
The numerical simulation based on the magnetic structure of the AFM3 phase can reproduce the observed NMR response.
We also find that the anisotropic hyperfine fields relate to the rotational symmetry breaking of the electronic structure in the AFM3 phase.
The relationship between the AFM3 phase and the electronic nematic state is still an important question that remains to be investigated.

\begin{acknowledgments}
We thank R. Kurihara for helpful discussions.
One of the authors (K. Matsui) is a Research Fellow of Japan Society for the Promotion of Science (JSPS).
This work was supported by the JSPS Grant-in-Aid for Scientific Research (Grant No. 18H01163, 19H01832).
\end{acknowledgments}

\end{document}